\documentstyle[epsf]{mn}

\newif\ifAMStwofonts
\AMStwofontstrue

\newcommand{\odn}[1]{(\protect\ref{#1})}
\newcommand{\ts}{\tau_{\rm T}}

\newcommand{\ntth}{\mbox{$\nu_{\rm t}^{\rm th}$}}
\newcommand{\nnth}{\mbox{$\nu_{\rm t}^{\rm nth}$}}
\newcommand{\xtth}{\mbox{$x_{\rm t}^{\rm th}$}}
\newcommand{\xnth}{\mbox{$x_{\rm t}^{\rm nth}$}}

\ifoldfss
  \ifCUPmtlplainloaded \else
    \NewTextAlphabet{textbfit} {cmbxti10} {}
    \NewTextAlphabet{textbfss} {cmssbx10} {}
    \NewMathAlphabet{mathbfit} {cmbxti10} {} 
    \NewMathAlphabet{mathbfss} {cmssbx10} {} 
  \fi
  \ifAMStwofonts
    \ifCUPmtlplainloaded \else
      \NewSymbolFont{upmath} {eurm10}
      \NewSymbolFont{AMSa} {msam10}
      \NewMathSymbol{\upi}     {0}{upmath}{19}
      \NewMathSymbol{\umu}     {0}{upmath}{16}
      \NewMathSymbol{\upartial}{0}{upmath}{40}
      \NewMathSymbol{\leqslant}{3}{AMSa}{36}
      \NewMathSymbol{\geqslant}{3}{AMSa}{3E}

       \let\le=\leqslant
       \let\ge=\geqslant
    \fi
  \fi
\fi 

\ifnfssone
  \newmathalphabet{\mathit}
  \addtoversion{normal}{\mathit}{cmr}{m}{it}
  \addtoversion{bold}{\mathit}{cmr}{bx}{it}
  \newmathalphabet{\mathbfit} 
  \addtoversion{normal}{\mathbfit}{cmr}{bx}{it}
  \addtoversion{bold}{\mathbfit}{cmr}{bx}{it}
  \newmathalphabet{\mathbfss} 
  \addtoversion{normal}{\mathbfss}{cmss}{bx}{n}
  \addtoversion{bold}{\mathbfss}{cmss}{bx}{n}
  \ifAMStwofonts
    \ifCUPmtlplainloaded \else
      \UseAMStwoboldmath
      \makeatletter
      \new@mathgroup\upmath@group
      \define@mathgroup\mv@normal\upmath@group{eur}{m}{n}
      \define@mathgroup\mv@bold\upmath@group{eur}{b}{n}
      \edef\UPM{\hexnumber\upmath@group}
      \new@mathgroup\amsa@group
      \define@mathgroup\mv@normal\amsa@group{msa}{m}{n}
      \define@mathgroup\mv@bold\amsa@group{msa}{m}{n}
      \edef\AMSa{\hexnumber\amsa@group}
      \makeatother
      \mathchardef\upi="0\UPM19
      \mathchardef\umu="0\UPM16
      \mathchardef\upartial="0\UPM40
      \mathchardef\leqslant="3\AMSa36
      \mathchardef\geqslant="3\AMSa3E

       \let\le=\leqslant
       \let\ge=\geqslant
    \fi
  \fi
\fi 

\ifnfsstwo
  \DeclareMathAlphabet{\mathbfit}{OT1}{cmr}{bx}{it}
  \SetMathAlphabet\mathbfit{bold}{OT1}{cmr}{bx}{it}
  \DeclareMathAlphabet{\mathbfss}{OT1}{cmss}{bx}{n}
  \SetMathAlphabet\mathbfss{bold}{OT1}{cmss}{bx}{n}
  \ifAMStwofonts
    \ifCUPmtlplainloaded \else
      \DeclareSymbolFont{UPM}{U}{eur}{m}{n}
      \SetSymbolFont{UPM}{bold}{U}{eur}{b}{n}
      \DeclareSymbolFont{AMSa}{U}{msa}{m}{n}
      \DeclareMathSymbol{\upi}{0}{UPM}{"19}
      \DeclareMathSymbol{\umu}{0}{UPM}{"16}
      \DeclareMathSymbol{\upartial}{0}{UPM}{"40}
      \DeclareMathSymbol{\leqslant}{3}{AMSa}{"36}
      \DeclareMathSymbol{\geqslant}{3}{AMSa}{"3E}

       \let\le=\leqslant
       \let\ge=\geqslant
    \fi
  \fi
\fi 

\ifCUPmtlplainloaded \else
  \ifAMStwofonts \else 
    \def\upi{\pi}
    \def\umu{\mu}
    \def\upartial{\partial}
  \fi
\fi

\title[Effects of non-thermal tails on synchrotron and Compton processes]
{Effects of non-thermal tails in Maxwellian electron distributions on
synchrotron and Compton processes}

\author[G. Wardzi\'nski and A. A. Zdziarski]
{{Grzegorz Wardzi\'nski\thanks{E-mail:
gwar@camk.edu.pl (GW), aaz@camk.edu.pl (AAZ)} and Andrzej A.
Zdziarski$^\star$}\\
N. Copernicus Astronomical Center, Bartycka 18, 00-716 Warszawa, Poland \\
}

\pagerange{\pageref{firstpage}--\pageref{lastpage}}
\pubyear{2000}

\begin{document}

\maketitle

\label{firstpage}

\begin{abstract} We investigate how the presence of a non-thermal tail beyond a
Maxwellian electron distribution affects the synchrotron process as well as
Comptonization in plasmas with parameters typical for accretion flows onto
black holes. We find that the presence of the tail can significantly increase
the net (after accounting for self-absorption) cyclo-synchrotron emission of
the plasma, which emission then provides seed photons for Compton upscattering.
Thus, the luminosity in the thermally-Comptonized spectrum is enhanced as well.
The importance of these effects increases with both increasing Eddington ratio
and the black hole mass. The enhancement of the Comptonized synchrotron
luminosity can be as large as by factors of $\sim 10^3$ and $\sim 10^5$ for
stellar and supermassive black holes, respectively, when the energy content in
the non-thermal tail is 1 per cent.

The presence of the tail only weakly hardens the thermal Comptonization
spectrum but it leads to formation of a high-energy tail beyond the thermal
cut-off, which two effects are independent of the nature of the seed photons.
Since observations of high-energy tails in Comptonization spectra can constrain
the non-thermal tails in the electron distribution and thus the Comptonized
synchrotron luminosity, they provide upper limits on the strength of magnetic
fields in accretion flows. In particular, the measurement of an MeV tail in the
hard state of Cyg X-1 by McConnell et al.\ implies the magnetic field strength
in this source to be at most an order of magnitude below equipartition.
\end{abstract}


\section{Introduction}
\label{s:introduction}

Power-law--like X-ray spectra extending up to soft $\gamma$-rays (hereafter
X$\gamma$)  are a common feature of the emission from accreting black holes in
active galactic nuclei (AGNs) and black-hole binaries (BHBs). It is generally
accepted that those spectra are produced by the Comptonization process taking
place in optically thin accreting plasmas with typical electron energies of
$\sim 100$ keV. Similar spectra are also observed from weakly-magnetized
accreting neutron stars (e.g.\ Barret et al.\ 2000), where they can originate
either in the accretion flow or in a boundary layer near the neutron star
surface.

The X$\gamma$ spectra of Seyfert 1s and X-ray binaries in their hard (low)
states can be modelled by Comptonization on thermal electrons, see e.g.\
Zdziarski (1999). However, both observations and theoretical considerations
suggest that a small non-thermal component may be present in the electron
distribution, giving rise to a weak non-thermal high-energy tail in the
observed Comptonization spectrum. Such a tail above $\sim 1$ MeV has been
detected  in the spectrum of Cyg X-1 in the hard state by the COMPTEL detector
aboard {\it CGRO\/} (McConnell et al.\ 1994, 2000a). In other BHBs and in
Seyferts, the existing upper limits are compatible with the presence of weak
non-thermal tails (e.g.\ Zdziarski et al.\ 1998; Gondek et al. 1996; Johnson et
al. 1997).

A non-thermal tail in the electron distribution in compact X$\gamma$ sources
can arise due to many different physical mechanisms. One such mechanism is
acceleration in the process of dissipation of magnetic field (similarly to the
case of the solar corona) in an optically-thin accretion flow or in active
coronal regions above an optically thick disc. Then, the energy loss and
subsequent thermalization of the accelerated electrons leads to a steady-state
electron distribution consisting of a Maxwellian and a non-thermal tail. Apart
from Coulomb interactions, an effective thermalizing process is synchrotron
self-absorption (Ghisellini, Guilbert \& Svensson 1988).

A strong support for the occurrence of non-thermal processes in black-hole
accretion flows comes from observations of X$\gamma$ spectra of black-hole
binaries in the soft state. In that case, Compton scattering by non-thermal
electrons appears to be the dominant process (Poutanen \& Coppi 1998;
Gierli\'nski et al.\ 1999). In the well-studied case of Cyg X-1, the
power-law--like spectrum from this process becomes dominant at $\ga 10$ keV
(Gierli\'nski et al.\ 1999) and it extends to at least $\sim 10$ MeV (McConnell
et al.\ 2000b).

An important issue concerns the source of soft photons undergoing
Comptonization in the accreting plasma. A natural candidate is thermal emission
from an optically thick accretion disc, either below the corona or outside the
region of optically thin accretion flow. However, as the presence of magnetic
fields of the order of that corresponding to equipartition with gas pressure
(i.e. $\sim 10^3$ G in AGNs and $\sim 10^7$ G in BHBs) is expected, one has to
take into account Comptonization of synchrotron photons (hereafter the CS
process) when calculating the emergent spectra. In models of
advection-dominated accretion flows, the CS process may be the main radiative
process in the innermost part of the accretion flow, where most of the
available gravitational energy is liberated (e.g.\ Narayan \& Yi 1995).

Wardzi\'nski \& Zdziarski (2000, hereafter WZ00) have investigated the
efficiency of the thermal CS process under physical conditions relevant for
accreting black holes and found that it can dominate only in low-luminosity
AGNs and in luminous stellar X-ray sources with the hardest spectra. In the
latter case, however, the presence of a correlation between the spectral index
and the strength of X-ray reflection (Zdziarski, Lubi\'nski \& Smith 1999;
Gilfanov, Revnivtsev \& Churazov 1999; Lubi\'nski \& Zdziarski 2001) appears to
imply that blackbody emission of cold media is the main source of seed photons
for thermal Comptonization, and the CS process is negligible.

In this work, we extend our previous results to the case of a hybrid,
quasi-thermal plasma and investigate the influence of a weak non-thermal
component in the electron distribution on the efficiency of the CS process. In
Section \ref{s:distributions}, we discuss the shape of the electron
distribution function resulting from acceleration processes operating in
accretion flows. The influence of non-thermal tails on the CS emission of the
plasma is discussed in Section \ref{s:emission}. We then investigate in detail
synchrotron emission and Comptonization spectrum of hybrid electrons in
Sections \ref{s:synchrotron} and \ref{s:comptonization}, respectively. This
allows us to estimate the efficiency of the CS process in hybrid plasmas
(Section \ref{s:cs}), and those results are applied to astrophysical black-hole
sources in Section \ref{s:applications}. Finally, we summarize our results in
Section \ref{s:discussion}.

\section{Hybrid electron distributions}
\label{s:distributions}

In accretion flows, hybrid electron distributions can be produced e.g.\
by stochastic acceleration of thermal electrons by interaction with plasma
instabilities, e.g.\ with whistlers via gyroresonance (as applied to the case
of Cyg X-1 by Li, Kusunose \& Liang 1996) or with fast-mode Alfv\'en waves
via transit-time damping (Li \& Miller 1997). Such process can result from
dissipation of magnetic field in an accretion disc corona (where magnetic
fields are raised from the disc by buoyant forces, Galeev, Rosner \& Vaiana
1979; Haardt, Maraschi \& Ghisellini 1994) or in an optically thin accretion
flow (Bisnovatyi-Kogan \& Lovelace 1997), provided a significant fraction of
the released energy generates turbulence.

Li \& Miller (1997) considered parameters typical for a plasma in an accretion
disc corona around a stellar black hole. They have shown, by solving the
Fokker-Planck equation, that electrons can be effectively accelerated from the
thermal background to relativistic energies, so that in the stationary state a
non-thermal tail of a power-law--like shape develops above certain Lorentz
gamma factor $\gamma_{\rm nth} \sim 2$ (which value is determined by the
efficiency of thermalization processes, see Section \ref{s:emission},) and the
power-law index, $p$, depends strongly on the plasma optical depth. The
power-law component is cut off at a $\gamma_{\rm f} \sim 10$ due to radiative
losses overcoming acceleration. Dermer, Miller \& Li (1996) predict somewhat
larger cut-off energies in the case of electron acceleration by whistlers and
in a plasma with parameters relevant for AGNs.

Similar shape of steady-state electron distribution is also predicted from
thermalization by synchrotron self-absorption and Coulomb interaction of an
injected power-law electron distribution (produced by any acceleration
mechanism), as shown by Ghisellini et al.\ (1988).

The exact shape of the non-thermal tail depends strongly on the plasma
parameters and the (uncertain) mechanism of magnetic field dissipation and
electron acceleration. We shall therefore consider a generic electron
distribution function of the form qualitatively reproducing that predicted by
detailed modelling of electron acceleration,
\begin{eqnarray}
\label{e:distribution}
\lefteqn{n_{\rm e}(\gamma) = \cases{n_{\rm e}^{\rm th}(\Theta, \gamma), &
$\gamma \le \gamma_{\rm nth}$;\cr
{n_{\rm e}^{\rm th}(\Theta, \gamma_{\rm
nth})\! \left(\frac{\gamma}{\gamma_{\rm nth}}\right)^{-p}\! \exp \!
\left({\gamma-\gamma_{\rm nth}\over \gamma_{\rm f}}\right)}, & $\gamma >
\gamma_{\rm nth}$.\cr} }
\end{eqnarray}
Here $\Theta\equiv kT/m_{\rm e} c^2$ is the dimensionless electron temperature
and $n_{\rm e}^{\rm th}(\Theta, \gamma)$ is the relativistic Maxwellian
normalized by the relation, $n_{\rm e}=\int_1^\infty n_{\rm e}(\gamma) {\rm d}
\gamma$, where $n_{\rm e}$ is the {\it total\/} electron density. This hybrid
electron distribution consists thus of a thermal part below $\gamma_{\rm nth}$
and an e-folded power-law above it (see also Section \ref{s:discussion}).

We note that the cases $p\la3$ and $p>3$ are qualitatively different. In the
former, most of the non-thermal synchrotron luminosity is produced by electrons
with $\gamma \la \gamma_{\rm f}$ and the non-thermal synchrotron spectrum is
hard. In the latter case, the luminosity is produced mostly by low-energy
electrons, $\gamma \sim \gamma_{\rm nth}$, and the spectrum is soft.
Observations of Cyg X-1 (Gierli\'nski et al.\ 1999; McConnell et
al.\ 2000a) suggest $p>3$. In the following we assume $p\ge3$.

\begin{figure}
\begin{center}
\noindent\epsfxsize=6.8cm\epsfbox{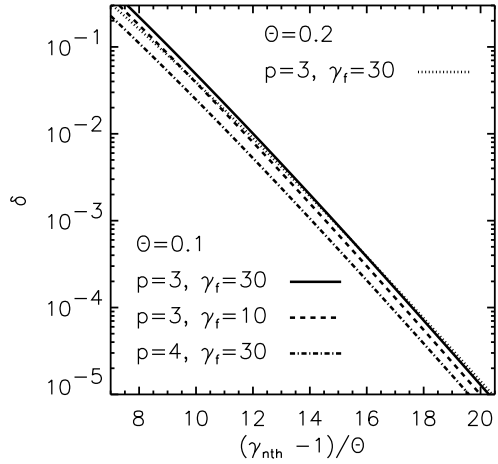}
\end{center}
\caption{The relation between $\gamma_{\rm nth}$ and $\delta$ for different
plasma parameters.
}
\label{f:delta}
\end{figure}

The degree to which a hybrid distribution is non-thermal can be measured by the
quantity, $(\gamma_{\rm nth}-1)/\Theta$, which shows how far in the Maxwellian
tail the non-thermal power law starts. Another such quantity is the ratio of
the energy densities in the non-thermal and thermal parts of the electron
distribution, $\delta$, which, for mildly relativistic temperatures and
$\gamma_{\rm nth}-1 \gg \Theta$ can be approximated as
\begin{eqnarray}
\label{e:eratio}
\lefteqn{\delta
\approx \frac{\int_{\gamma_{\rm nth}}^\infty (\gamma -1)n_{\rm
e}(\gamma) {\rm d}\gamma}{\frac{6+15\Theta}{4+5\Theta} \Theta n_{\rm e}} }\\
\nonumber \lefteqn{
\approx \frac{4+5\Theta}{6+15\Theta}
\frac{\left(\gamma_{\rm nth}^2-1\right)^{1/2}}{\Theta^2
K_2\left({1\over\Theta}\right)}
\left(\frac{\gamma_{\rm nth}^3}{p-2}-\frac{\gamma_{\rm nth}^2}{p-1}\right){\rm
e}^{-\frac{\gamma_{\rm nth}}{\Theta}}  ,}
\end{eqnarray}
where $K_{2}$ is a modified Bessel function and we employed an approximation
for the thermal energy density from Gammie \& Popham (1998). In the above
derivation (as well as in all analytical approximations below) we neglected,
for simplicity, the exponential cut-off in the distribution
\odn{e:distribution}. Typical relations between $\delta$ and $(\gamma_{\rm
nth}-1)/\Theta$ are shown in Fig.\ \ref{f:delta}.

Since the exponential term in equation \odn{e:eratio} varies fastest,
we have an approximate dependence,
\begin{equation}
\label{e:handwaving}
\gamma_{\rm nth} \approx {C}-\Theta \ln \delta,
\end{equation}
where $C\sim 1$ is slowly increasing with increasing $\Theta$ and $\gamma_{\rm
nth}$. We then see that $(\gamma_{\rm nth}-1)/\Theta$ only weakly depends on
$\Theta$ and the shape of the tail for a given $\delta$, see Fig.\
\ref{f:delta}.

\section{General form of emission from a hybrid electron distribution}
\label{s:emission}

In accreting black holes, synchrotron emission from thermal plasmas optically
thin to scattering is usually strongly self-absorbed up to a turnover
frequency, $\nu_{\rm t}$, below which the plasma is optically thick to
absorption,
\begin{equation}
\label{e:turnover}
\alpha_{\nu_{\rm t}}R=1,
\end{equation}
where $\alpha_{\nu}$ is the absorption coefficient and $R$ is the
characteristic
size of the source. Hereafter, we assume that $\nu_{\rm t}$ is determined only
by synchrotron absorption, i.e. other radiative processes are negligible at
$\nu \sim \nu_{\rm t}$. Typically, accretion flow models predict (assuming
equipartition magnetic fields) $\nu_{\rm t} \sim 10^{15}$ Hz for BHBs and
$\nu_{\rm t}\sim 10^{12}$ Hz for AGNs, see e.g. WZ00. This corresponds
to the value of $\nu_{\rm t}/\nu_{\rm c}$ from tens to hundreds, where
$\nu_{\rm c}=eB/2\pi m_{\rm e} c$ is the cyclotron frequency for the magnetic
field of the strength $B$ and $m_{\rm e}$ and $c$ are, respectively, the rest
mass of the electron and the velocity of light. The ratio, $\nu_{\rm
t}/\nu_{\rm c}$, is larger in AGNs as a result of their magnetic field weaker
than that in BHBs. The Lorentz factor, $\gamma_{\rm t}$, of the electrons that
emit most of their synchrotron radiation at $\nu_{\rm t}$ is of the order of
$\gamma_{\rm t} \sim (\nu_{\rm t}/\nu_{\rm c})^{1/2}$. The value of
$\gamma_{\rm t}$ is of the order of a few, again larger in the case of
AGNs than of BHBs.

First, let us consider a homogenous cloud of thermal plasma. The CS spectrum
below the thermal turnover frequency, $\ntth$, can be approximated as a
Rayleigh-Jeans spectrum. Above  $\ntth$, the synchrotron emission drops rapidly
and the spectrum is often dominated by a power-law component resulting from
Comptonization, which is cut off above an energy of $\sim kT$, where $T$ is the
electron temperature and $k$ is the Boltzmann constant.

\begin{figure}
\begin{center}
\noindent\epsfxsize=6.8cm\epsfbox{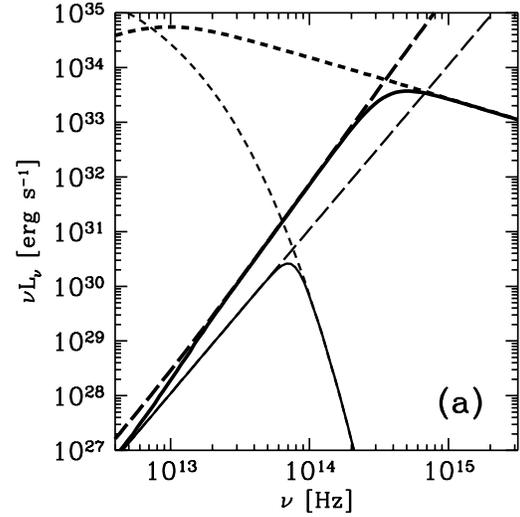}
\end{center}
\nobreak
\begin{center}
\noindent\epsfxsize=6.8cm\epsfbox{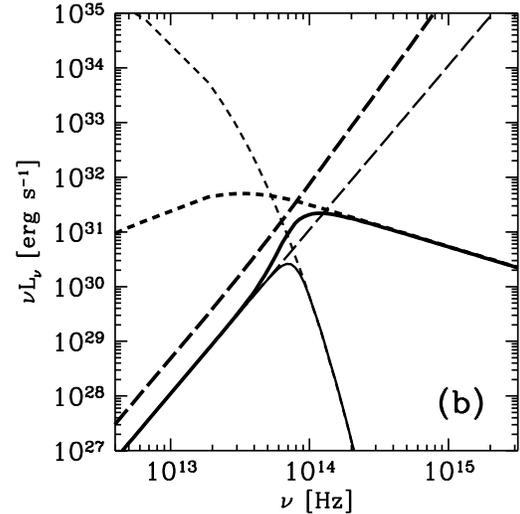}
\end{center}
\caption{Synchrotron spectra from a spherical source with the energy content of
the non-thermal tail of (a) $\delta=0.19$, (b) $\delta=10^{-4}$, and other
plasma parameters as in our model of emission from Cyg X-1 in Section
\protect\ref{s:applications}. The heavy short-dashed curve shows the
optically-thin emission of the non-thermal electrons alone, $(4/3)\upi
R^3(4\upi j_{\nu}^{\rm pl})$, where $j_{\nu}^{\rm pl}$ is their emission
coefficient. The heavy long-dashed curve shows the self-absorbed emission of
the non-thermal electrons, $4\upi^2R^2S_{\nu}^{\rm pl}$, where $S_{\nu}^{\rm
pl}$ is their source function. The thin short-dashed and long-dashed cuves show
the thermal ($\delta=0$) optically-thin and optically-thick emission,
respectively. The heavy and thin solid curves give the corresponding emergent
spectra, which change  from optically thick to optically thin around the
turnover frequency.
}
\label{f:widma}
\end{figure}

Now, let us assume that the electron distribution is given by equation
\odn{e:distribution}. If $\gamma_{\rm nth} < \gamma_{\rm t}$, the number of
electrons with $\gamma \sim \gamma_{\rm t}$ will increase (compared to the
thermal case) as well as the absorption coefficient and in consequence the new
turnover frequency, $\nnth$, will be larger. Also the shape of the synchrotron
spectrum is modified in this case. At low frequencies  both emission and
absorption are dominated by thermal electrons so that the optically thick
spectrum is still the Rayleigh-Jeans one. However, at higher frequencies the
emission of non-thermal electrons starts to dominate, while absorption is still
dominated by thermal electrons [which is a consequence of the thermal
distribution falling off much more steeply than the non-thermal one, see
equation \odn{e:alpha}], and the optically-thick spectrum quickly increases.
This leads to the regime in which both the non-thermal emission and absorption
dominate, in which case the spectrum is proportional to the source function of
non-thermal electrons. Finally, the spectrum  becomes optically thin ($\propto
\nu^{-(p-1)/2}$), turning over above $\nnth$.

This behaviour is shown in Fig. \ref{f:widma} where the resulting synchrotron
spectra are calculated for two different energy contents of the non-thermal
tail and, as a reference, for the purely thermal case. Note that the
non-thermal tail has to be sufficiently strong for the optically thick spectrum
to reach the non-thermal source function before it turns over.

Both optically thick and optically thin parts of the synchrotron spectrum will
undergo Comptonization. This process will thus be modified with respect to the
thermal case due to two effects. Firstly, Compton scattering itself will
involve both thermal electrons as well as non-thermal ones. The effect of
non-thermal electrons on Comptonization will be most prominent at energies
$h\nu \gg kT$ (where $h$ is the Planck constant), where a high-energy tail
(above the thermal cut-off) will develop. Secondly, the seed-photon flux will
be higher, so luminosity in the thermal Comptonization spectrum will also
increase. Since $\gamma_{\rm t}$ in optically-thin plasmas around black holes
is expected to be well above the mean energy of the thermal electrons, even a
non-thermal tail carrying a tiny fraction of the total electron energy, and
thus only weakly modifying the shape of the Comptonization spectrum, can
increase the turnover frequency and thus the luminosity produced by the CS
process by a large factor.

Then a question arises whether the presence of strong synchrotron
self-absorption, acting as a thermalizing mechanism for electrons of $\gamma
\la \gamma_{\rm t}$ (i.e. in the regime of optically thick emission), will
allow for an electron distribution deviating substantially from a pure
Maxwellian for $\gamma \sim \gamma_{\rm t}$. This issue was addressed by
Ghisellini et al.\ (1988), who have shown that an initial power-law
distribution of injected electrons will develop (after the thermalization time
scale equal to a few synchrotron cooling timescales, $t_{\rm c}^{\rm syn}$,
defined for the mean energy of the electrons radiating in the optically thick
range of the spectrum) into a Maxwellian stationary distribution for $\gamma <
\gamma_{\rm t}$, provided the synchrotron cooling dominates and $t_{\rm c}^{\rm
syn} \ll t_{\rm esc}$ (where $t_{\rm esc}$ is the escape time from the source).
Assuming equipartition of magnetic field pressure with gas pressure, we find
the latter criterion is always fulfilled in accretion flows onto black holes,
independent of the form of the injected distribution.

However, when the inverse Compton process is dominant as a cooling mechanism,
both $\gamma_{\rm nth}$ and $kT$ diminish and a power-law tail starting below
$\gamma_{\rm t}$ is formed, see Ghisellini \& Svensson (1990). A sufficient
condition for the dominance of Compton cooling is $\alpha<1$, which is
typically fulfilled in astrophysical accreting black holes. Even when
$\alpha>1$, Comptonization of external soft photons (e.g.\ from a cold
accretion disk) can dominate electron cooling (see, e.g.\ Ghisellini et al.\
1998). Thus, $\gamma_{\rm nth}<\gamma_{\rm t}$ is a likely condition in those
sources.

\section{Synchrotron turnover frequency}
\label{s:synchrotron}

Let $\bar{\eta}_\nu(\gamma)$ be the synchrotron emission coefficient of a
single electron averaged both over the electron velocity direction and the
direction of emission. Then, the absorption coefficient for an isotropic
distribution of electrons in a chaotic magnetic field is (e.g.\ Ghisellini \&
Svensson 1991) is
\begin{equation}
\label{e:alpha}
\alpha_\nu= \frac{-1}{2 m_{\rm e}\nu^2} \int\limits_{1}^\infty \gamma
\left(\gamma^2-1\right)^\frac{1}{2}
\bar{\eta}_\nu(\gamma)
\frac{\rm d}{{\rm d} \gamma}\left[\frac{n_{\rm e}(\gamma) }{\gamma
(\gamma^2-1)^\frac{1}{2} }\right] {\rm d} \gamma.
\end{equation}

We assume a homogenous spherical source of the radius, $R$, and the Thomson
optical depth, $\tau_{\rm T}$, where $\tau_{\rm T}\equiv n_{\rm e}\sigma_{\rm
T}R$ and $\sigma_{\rm T}$ is the Thomson cross section. Then, we can
numerically calculate the absorption coefficient with equation \odn{e:alpha}
for the distribution of equation \odn{e:distribution} and then solve equation
\odn{e:turnover} for the turnover frequency, $\nnth$. This can be compared with
the {\it thermal\/} turnover frequency, which for plasma parameters typical for
accreting black holes can be approximated as (WZ00)
\begin{equation}\label{e:nuthapp}
\nu_{\rm t}^{\rm th} \approx 7\times 10^{3} \Theta^{0.95}
\ts^{0.05} \nu_{\rm c}^{0.91} \; {\rm Hz}.
\end{equation}

On the other hand, we can obtain an analytical estimate, $\nu_{\rm t}^{\rm
pl}$, of the turnover frequency for the hybrid electron distribution by
considering a purely power-law electron distribution normalized to match the
Maxwellian at $\gamma_{\rm nth}$. For the synchrotron absorption coefficient
for power-law electrons (Rybicki \& Lightman 1979),
\begin{equation}
\label{e:nupl}
\nu_{\rm t}^{\rm
pl}=3^{\frac{1+p}{4+p}}2^{-\frac{6}{4+p}}\upi^{\frac{1}{4+p}}\nu_{\rm
c}^{\frac{2+p}{4+p}}\left[G_1 R r_{\rm e} c n_{\rm e}(\gamma_{\rm nth})
\gamma_{\rm nth}^p \right]^{\frac{2}{4+p}},
\end{equation}
which, with equation \odn{e:eratio}, leads to
\begin{eqnarray}
\label{e:nupl2}
\lefteqn{\nu_{\rm t}^{\rm
pl}\approx3^{\frac{1+p}{4+p}}2^{-\frac{6}{4+p}}\upi^{\frac{1}{4+p}}\nu_{\rm
c}^{\frac{2+p}{4+p}}\left(\frac{\gamma_{\rm
nth}}{p-2}-\frac{1}{p-1}\right)^{-\frac{2}{4+p}}}\nonumber\\
\lefteqn{\times \left(
\frac{6+15\Theta}{4+5\Theta}\frac{G_1 r_{\rm e} c \tau_{\rm T} \delta \Theta
\gamma_{\rm nth}^{p-1}}{\sigma_{\rm T}}
\right)^{\frac{2}{4+p}},}
\end{eqnarray}
where
\begin{equation}
\label{e:g1}
G_1
=\frac{\Gamma(\frac{6+p}{4})\Gamma(\frac{2+3p}{12})\Gamma(\frac{22+3p}{12})}
{\Gamma(\frac{8+p}{4})} \simeq 1,
\end{equation}
$r_{\rm e}$ is the classical electron radius and $\Gamma$ is Euler's gamma
function.

Provided $p \ga 3$ (so that radiation from electrons with $\gamma > \gamma_{\rm
f}$ is negligible) and emission at $\nnth$ is dominated by non-thermal
electrons (i.e. $\gamma_{\rm f} > \gamma_{\rm t} > \gamma_{\rm nth}$, which
condition can be checked a posteriori by comparing the source function of
power-law electrons to the Rayleigh-Jeans intensity), $\nnth\simeq \nu_{\rm
t}^{\rm pl}$. On the other hand, $\nu_{\rm t}^{\rm pl}<\nu_{\rm t}^{\rm th}$
when emission from thermal electrons is significant at $\nnth$ and, in general,
\begin{equation}
\label{e:maxnu} \nnth\simeq {\rm max\left(\ntth, \nu_{\rm t}^{\rm
pl}\right)}.
\end{equation}
The accuracy of this approximation is typically $\la 10$ per cent
(provided $\gamma_{\rm t}<\gamma_{\rm f}$ and $p \ga 3$). See
WZ00 for a discussion of approximations of $\ntth$.

We can obtain the dependences of $\nu_{\rm t}^{\rm nth}/\ntth$ on the plasma
parameters by considering the ratio $\nu_{\rm t}^{\rm pl}/\ntth$ [given by
approximations \odn{e:nuthapp} and \odn{e:nupl}]. First, at constant $\delta$,
this ratio decreases with increasing $\Theta$ since then $\gamma_{\rm nth}$
increases with $\Theta$ [see equation \odn{e:handwaving}]. The dependence on
$\tau_{\rm T}$ is stronger for $\nu_{\rm t}^{\rm pl}$ than for $\ntth$ as a
result of much steeper optically thin synchrotron spectrum in the latter case.
This leads to $\nu_{\rm t}^{\rm pl}/\ntth \propto  \tau_{\rm
T}^{2/(4+p)-0.05}$, i.e. the larger the optical depth, the larger the ratio
$\nu_{\rm t}^{\rm pl}/\ntth$. In a hot accretion flow, we generally expect
$\tau_{\rm T}$ increasing and $\Theta$ decreasing with increasing accretion
rate, so the effect of a non-thermal tail (of a given form) on the value of the
turnover frequency can be expected to be the largest for the most luminous
sources.

Since $\nu_{\rm t}^{\rm pl}/\ntth \propto \nu_{\rm c}^{(2+p)/(4+p)-0.91}$, the
non-thermal increase of the turnover frequency grows with decreasing
$\nu_{\rm c}$. This effect is due to the increase of $\gamma_{\rm t}$ with
decreasing $\nu_{\rm c}$ [as a result of increasing $\ntth/\nu_{\rm c}$, see
equation \odn{e:nuthapp}], which leads to a much larger ratio of the number of
non-thermal to thermal electrons at $\gamma_{\rm t}$ for a given  $\gamma_{\rm
nth}$. Thus, the effect of a non-thermal tail with a given form will be much
stronger in AGNs than in BHBs (due to much weaker magnetic field in the
former objects).

\begin{figure}
\begin{center}
\noindent\epsfxsize=6.7cm\epsfbox{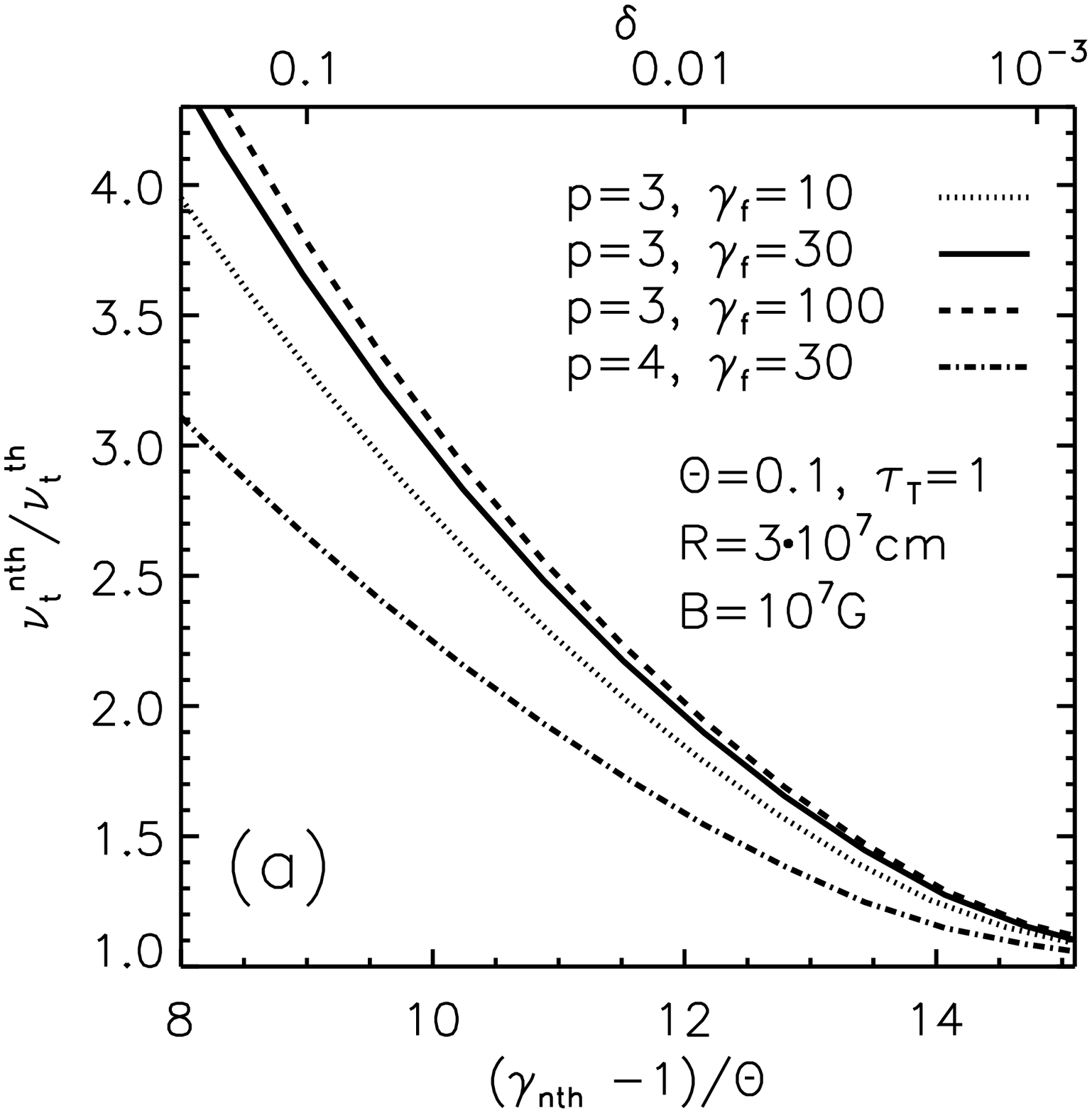}
\end{center}
\nobreak
\begin{center}
\noindent\epsfxsize=6.7cm \epsfbox{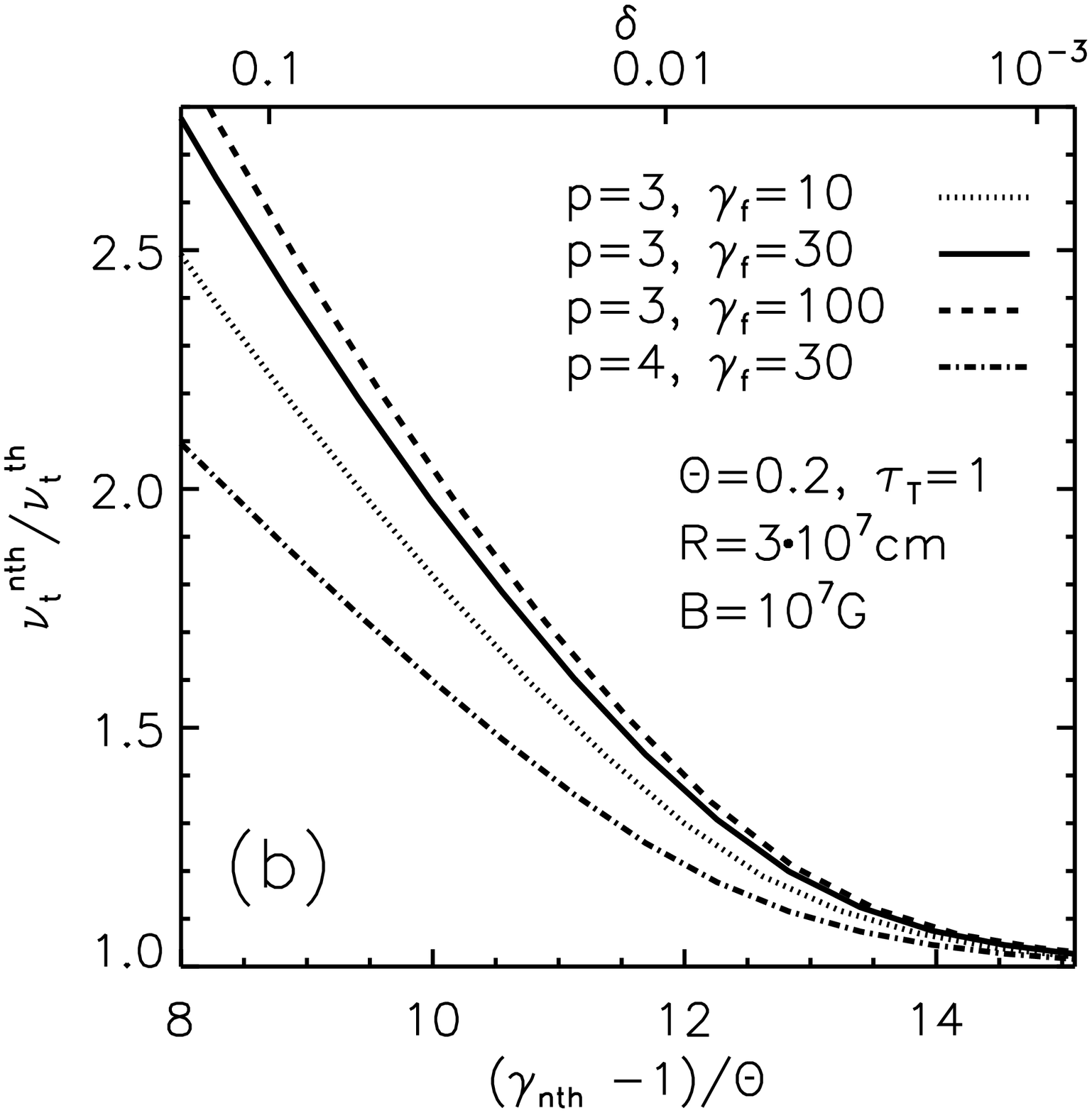}
\end{center}
\begin{center}
\noindent\epsfxsize=6.7cm \epsfbox{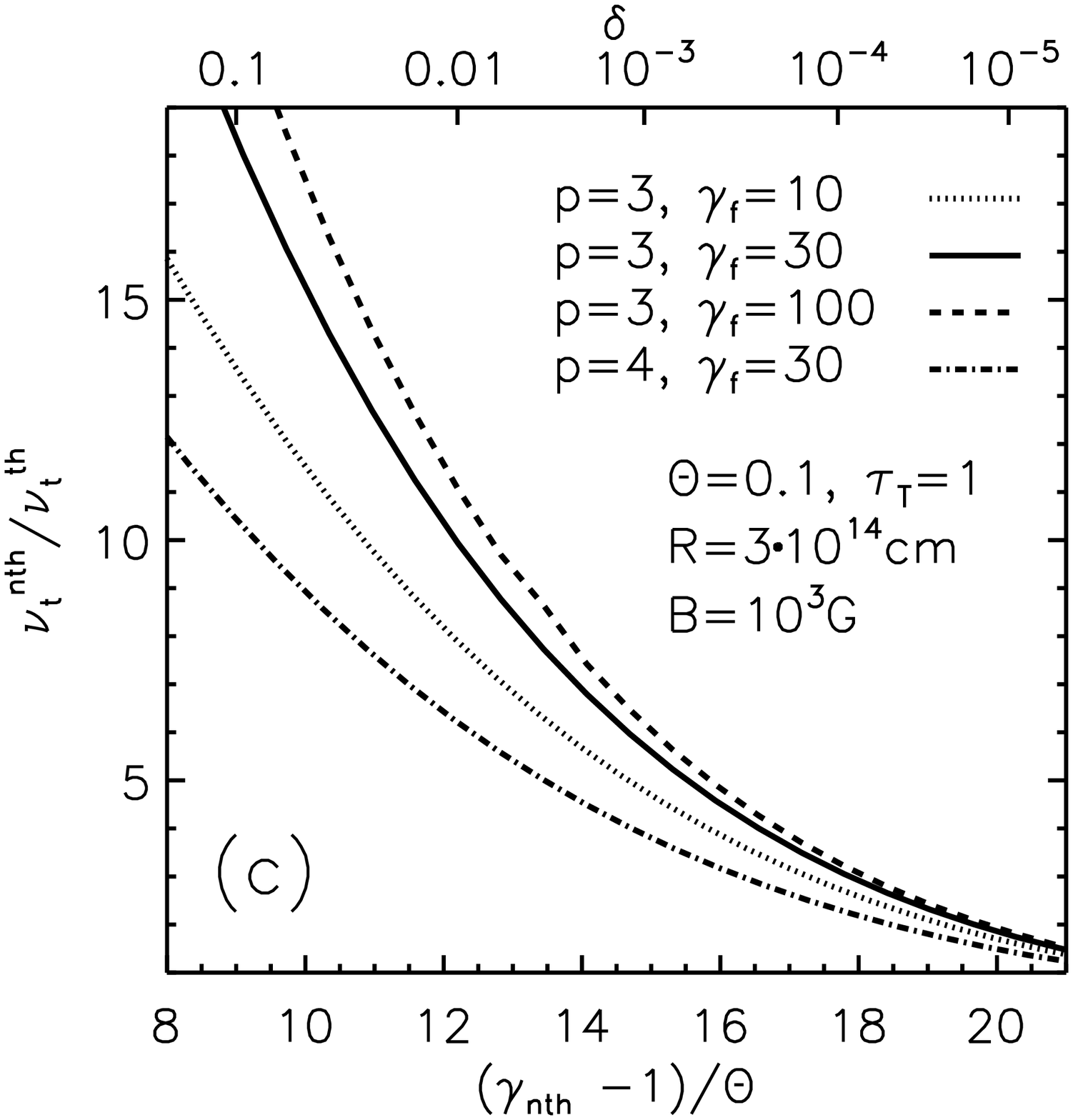}
\end{center}
\caption{The increase of $\nu_{\rm t}$ as a function of the electron
energy where the non-thermal tail starts for
various plasma parameters and forms of the tail. The upper axis
shows the values of $\delta$ for $p=3$ and $\gamma_{\rm f}=30$. Note that the
relation between $(\gamma_{\rm nth}-1)/\Theta$ and $\delta$ is slightly
different for the curves with other values of $p$ and $\gamma_{\rm f}$, see
Fig.\ \protect\ref{f:delta}.}
\label{f:ratio}
\end{figure}

Fig.\ \ref{f:ratio} shows the values of $\nnth/\ntth$ as a function of
$(\gamma_{\rm nth}-1)/\Theta$ for three sets of thermal plasma parameters. In
the first and second set, we assume parameters relevant for luminous BHBs,
$\tau_{\rm T}$=1, $R=3 \times 10^7$ cm, $B=10^{7}$ G, and $\Theta=0.1$ and 0.2,
respectively. In the third set, we assume parameters relevant to AGNs,
$\tau_{\rm T}$=1, $R=3\times 10^{14}$ cm, $B=10^{3}$ G and  $\Theta=0.1$. For
each set, we consider two different slopes of the non-thermal tail, $p=3$ and 4
and $\gamma_{\rm f}=30$. Additionally, we consider the cases of $\gamma_{\rm
f}=10$ and 100 for $p=3$ (for $p>3$ the influence of the cut-off is much
weaker). We see that even a weak non-thermal component, with $(\gamma_{\rm
nth}-1)/\Theta \sim 12$ (which corresponds to only $\sim 1$ per cent of the
total energy density of the electrons in the non-thermal tail) can lead to an
increase of the turnover frequency by a factor of $\sim 1.5$--2 for BHBs and
$\sim 10$ for AGNs.

The dependence of the relative increase of the turnover frequency on $p$ and
$\gamma_{\rm f}$ is relatively weak. This reflects the fact that the turnover
frequency depends mostly on the number of electrons at $\gamma_{\rm t}$ and
since for small values of $\delta$, $\gamma_{\rm t}$ is only slightly larger
than $\gamma_{\rm nth}$, $n_{\rm e}(\gamma_{\rm t})$ depends weakly on either
$p$ or $\gamma_{\rm f}$. This suggests that the degree of the dependence on $p$
and $\gamma_{\rm f}$ should increase with increasing $\nnth/\ntth$, in
agreement with our results. It is important to note that weak dependence of
$\nnth/\ntth$ on the shape of the non-thermal tail makes our results weakly
dependent on details of the acceleration mechanism.

\section{Comptonization by a hybrid electron distribution}

\subsection{The general case}
\label{s:comptonization}

For Comptonization by a thermal plasma, we employ here a simple treatment of
this process of Zdziarski (1985, 1986). In that approximation, the
Comptonization spectrum above the frequency, $\nu_{\rm inj}$, at which soft
seed photons are injected (which, in the case of the CS process, $\approx
\ntth$), can be approximated as a sum of an e-folded power-law with energy
index $\alpha$ and a Wien spectrum,
\begin{equation}
\label{e:lthx}
\frac{{\rm d} L_{\rm C}^{\rm th}}{{\rm d} x} \propto \left[ \left({x \over
\Theta}\right)^{-\alpha}
+ \frac{\Gamma(\alpha)P_{\rm sc}}{\Gamma(2\alpha+3)}\left({x \over
\Theta}\right)^{3} \right]{\rm e}^{-x/\Theta},
\end{equation}
where  $x\equiv h\nu/m_{\rm e}c^2$ (hereafter all indices of $x$ have the same
meaning as those of $\nu$) and $P_{\rm sc}$ is the volume-averaged
scattering probability. Its integration yields the thermal-Compton luminosity,
which, in the case of a spherical source, is
\begin{equation}
\label{e:lth}
L_{\rm C}^{\rm th}({\cal C}) \simeq 4 \upi R^2 {\cal C} \Theta
\left[\Gamma\left(1-\alpha,{x_{\rm inj} \over
\Theta}\right)+\frac{6\Gamma(\alpha)P_{\rm sc}}{\Gamma(2\alpha+3)}\right],
\end{equation}
where ${\cal C}$ is a constant depending on the flux  of the injected seed
photons (we calculate ${\cal C}$ in Section \ref{s:cs} for the case of
synchrotron seed photons).

The presence of non-thermal electrons modifies the Comptonization spectrum and
since for the electron distributions we consider (i.e. $\tau_{\rm T}\sim 1$
and $\delta \ll 1$) the Thomson optical depth for scattering off non-thermal
electrons is $\ll 1$, the resulting spectrum can be approximated as a
convolution of the thermal Comptonization spectrum with a spectrum resulting
from single scattering of photons off non-thermal electrons. Therefore, for $x
\la \Theta$, the Comptonization spectrum becomes harder, while for $x \gg
\Theta$, above the thermal cut-off, a power-law tail in the spectrum develops.
As a result, the overall luminosity produced by Comptonization increases.

\begin{figure}
\begin{center}
\noindent\epsfxsize=6.8cm \epsfbox{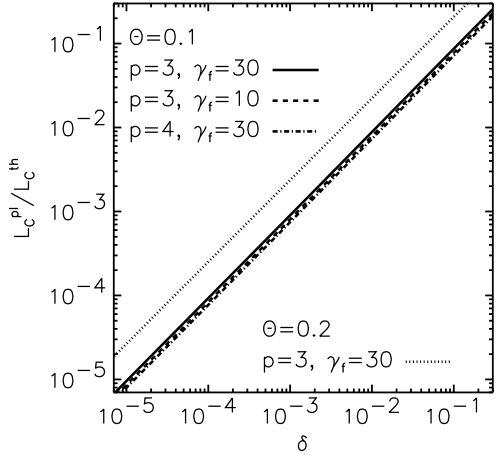}
\end{center}
\caption{
The relation between the fraction of energy carried by non-thermal electrons
and the ratio of $L_{\rm C}^{\rm pl}/L_{\rm C}^{\rm th}$ for different plasma
parameters and $\alpha=0.7$.
}
\label{f:compare}
\end{figure}

In calculations of the luminosity from Comptonization on non-thermal electrons,
$L_{\rm C}^{\rm pl}$, where the Klein-Nishina effect has to be taken into
account, we use the approximation for the rate of energy change, ${\rm d}\gamma
/{\rm d}t$, of a single electron via Compton interaction with isotropic photons
of energy density, $U_{\rm ph}$, and mean energy, $\langle x\rangle$,
\begin{equation}
\label{e:loss}
\frac{{\rm d} \gamma}{{\rm d} t}=-{4 \over 3}{\sigma_{\rm T}
\left(\gamma^2-1\right) U_{\rm ph} \over m_e c} \left[
1-\frac{63}{10}\frac{\gamma\langle x^2\rangle}{\langle x\rangle} \right],
\end{equation}
where the Klein-Nishina cross-section was approximated using the first-order
correction to the Thomson-limit cross-section (Rybicki \& Lightman 1979).
We assume that the photons undergoing scattering off non-thermal electrons are
those from the thermal Comptonization spectrum above $x_{\rm inj}$. Then
\begin{equation}
\label{e:flux}
U_{\rm ph}(x)=\frac{3 \left({\rm d}L_{\rm C}^{\rm th}/{{\rm d}
x}\right)\left({3/4}+{\tau_{\rm T}/5}\right)}{4 \upi c
R^2},
\end{equation}
where the factor $\left(3/4+\tau_{\rm T}/5\right)$ accounts for the change of
the escape time due to scatterings in the source and is a matching formula
between the optically thin case (where the escape time is $3R/4c$) and the
optically thick one ($\tau_{\rm T} R/5c$, Sunyaev \& Titarchuk 1980). We can
further simplify the calculations assuming that the spectrum of photons
undergoing non-thermal Comptonization is a pure power-law and then neglect
scatterings in the Klein-Nishina limit (see below).

Now, for each $\gamma$ we calculate $U_{\rm ph}$, $\langle x\rangle$ and
$\langle x^2\rangle$ as an integral over the power-law spectrum from $x_{\rm
inj}$ up to the energy, $x_{\rm max}$, of the limit of the Klein-Nishina
regime, which we obtain from the condition of $1-63/(10\gamma x_{\rm max})=0$,
in which the numerical coefficient is chosen for consistency with equation
(\ref{e:loss}).

Then the formula \odn{e:loss} is integrated over $n_{\rm e}(\gamma)$, from
$\gamma_{\rm nth}$ to infinity, which leads to
\begin{eqnarray}
\label{e:lnth}
\lefteqn{L_{\rm C}^{\rm pl}({\cal C})=\frac{16 \upi}{3} \left({10 \over
63}\right)^{1-\alpha} \frac{6+15\Theta}{4+5\Theta} \left({3 \over 4}+{\tau_{\rm
T}\over 5}\right)
\frac{R^2 {\cal C} \delta \tau_{\rm T}\Theta^{\alpha+1}
}{(1-\alpha)(2-\alpha)}}
\nonumber\\
\lefteqn{\times
\left(\frac{\gamma_{\rm nth}}{p-2}-\frac{1}{p-1}\right)^{-1}
\left(\frac{\gamma_{\rm nth}^{\alpha+1}}{p-\alpha-2}-\frac{\gamma_{\rm
nth}^{\alpha-1}}{p-\alpha}\right) ,}
\end{eqnarray}
where equation \odn{e:eratio} has been employed.

The total luminosity, $L_{\rm C}^{\rm nth}=L_{\rm C}^{\rm th}+L_{\rm C}^{\rm
pl}$, as well as the ratio $L_{\rm C}^{\rm pl}/L_{\rm C}^{\rm th}$ can now be
directly obtained from expressions \odn{e:lth} and \odn{e:lnth}.
In the range of $0.4\la \alpha\la 0.9$, both the high-energy part of the power
law component dominates the luminosity as well as the Wien component can still
be neglected. Then, we obtain (see WZ00)
\begin{eqnarray}
\label{e:powrat}
\lefteqn{\frac{L_{\rm C}^{\rm pl}}{L_{\rm C}^{\rm th}} \approx \frac{4}{3}
\left({10 \over 63}\right)^{1-\alpha}\frac{6+15\Theta}{4+5\Theta} \left({3
\over 4}+{\tau_{\rm T}\over
5}\right)\frac{\delta \tau_{\rm T} \Theta^{\alpha}}{2-\alpha}  } \nonumber\\
\lefteqn{\times
\left(\frac{\gamma_{\rm nth}}{p-2}-\frac{1}{p-1}\right)^{-1}
\left(\frac{\gamma_{\rm nth}^{\alpha+1}}{p-\alpha-2}-\frac{\gamma_{\rm
nth}^{\alpha-1}}{p-\alpha}\right)
.}
\end{eqnarray}
We find the accuracy of this approximation is typically
$\la 30$ per cent. Figure \ref{f:compare} shows that while the ratio $L_{\rm
C}^{\rm pl}/L_{\rm
C}^{\rm th}$ depends strongly on
$\delta$, the dependence on the non-thermal tail shape is very weak.

\begin{figure}
\begin{center}
\noindent\epsfxsize=8.4cm\epsfbox{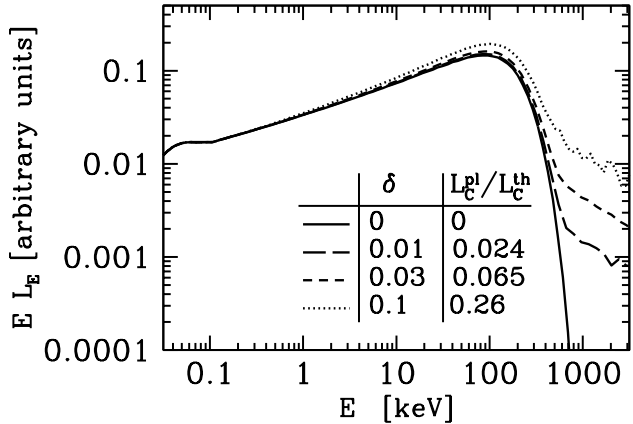}
\end{center}
\caption{Monte Carlo Comptonization spectra of 10-eV blackbody photons in a
plasma of $\Theta=0.1$, $\tau_{\rm T}=2.4$. The solid line represents the
thermal spectrum, while the other lines correspond to the cases of power-law
tails in the electron distribution of $p=3$, $\gamma_{\rm f}=30$ and different
values of $\delta$. }
\label{f:spectra}
\end{figure}

In Fig.\ \ref{f:spectra}, we show as an example Comptonization spectra from a
homogenous plasma cloud, produced with the Monte Carlo code of Gierli\'nski
(2000). We see that as long as $\delta < 0.1$ the Comptonization spectrum at $x
\la \Theta$ is hardly modified by the presence of non-thermal electrons and the
onset of the power-law tail is no higher than an order of magnitude below the
peak in the $xL(x)$ spectrum.

From the observational point of view, an important feature measuring the
deviation from purely thermal Comptonization spectrum is the fraction of the
luminosity radiated in the high-energy tail beyond the thermal spectrum. The
ratio $L_{\rm C}^{\rm pl}/L_{\rm C}^{\rm th}$ is larger than this fraction as
it incorporates also the non-thermal luminosity at energies below the tail.
This
effect is stronger for smaller $\delta$, when a larger fraction of $L_{\rm
C}^{\rm pl}$ is hidden below the thermal peak. For example, in the spectra in
Fig.\ \ref{f:spectra} corresponding to $\delta=0.1$ and $\delta=0.01$ the
high-energy tail develops at $\sim 300$ keV and $\sim 500$ keV, respectively.
Then $L_{\rm C}^{\rm pl}/L_{\rm C}^{\rm th}$ is larger than he fraction of the
luminosity in the tail by factors 3 and 5, respectively. Thus, $L_{\rm C}^{\rm
pl}/L_{\rm C}^{\rm th}$ provides only an upper limit to the
fraction of the luminosity radiated in the high-energy tail.

\subsection{Comptonization of synchrotron photons}
\label{s:cs}

As we have seen above, the presence of a non-thermal component in the thermal
distribution of electrons in a plasma where synchrotron radiation is produced
and then Comptonized leads to an increase of the luminosity available from
this process in two ways. First, it increases the number of soft photons (both
from optically thick emission up to $\xnth$ and from optically
thin emission above $\xnth$) and second, it increases the efficiency of
Comptonization, the former effect being significantly stronger.

In the case of purely thermal synchrotron Comptonization, the
normalization of the Comptonization spectrum can be assumed (Zdziarski 1985,
1986) to be proportional to the Rayleigh-Jeans flux at the turnover frequency,
so that the constant ${\cal C}$ in the expression
\odn{e:lth}, which we now denote as ${\cal C}^{\rm th}$, reads,
\begin{equation}
\label{e:cth}
{\cal C}^{\rm th}=\frac{2\upi m_{\rm e}c^3}{\lambda_{\rm
C}^3}\varphi\Theta^{1-\alpha}\left(\xtth\right)^{2+\alpha},
\end{equation}
where
\begin{equation}
\label{e:varphi}
\varphi(\Theta)\approx
\frac{1+ (2\Theta)^2}{1+ 10 (2 \Theta)^2}
\end{equation}
is a heuristic formula obtained by Zdziarski (1985) to match Monte Carlo
simulation, and $\lambda_{\rm C}$ is the Compton wavelength.
The total luminosity from the CS process is then
\begin{equation}
\label{e:lcsth}
L_{\rm CS}^{\rm th}=L_{\rm S}^{\rm th}+L_{\rm C}^{\rm th}({\cal C}^{\rm th}),
\end{equation}
where $L_{\rm S}^{\rm th}$ is the luminosity in the thermal synchrotron
spectrum.

If thermal and hybrid synchrotron photons experience the same average number of
scatterings off thermal electrons before reaching $x\sim\Theta$ (i.e.\ when
$\xnth/\xtth \ll \Theta/\xtth$ and the synchrotron spectrum is sufficiently
narrow, which corresponds to $p>3$ and which we assume hereafter in this
section),
the ratio of the luminosity from Comptonization off thermal electrons in the
hybrid case to that in the thermal case will equal the corresponding ratio of
the synchrotron luminosities. Then,
\begin{equation}
\label{e:softflux}
\frac{{\cal C}^{\rm nth}}{{\cal C}^{\rm th}} = \frac{L_{\rm S}^{\rm
nth}}{L_{\rm S}^{\rm th}},
\end{equation}
where $L_{\rm S}^{\rm nth}$ is the hybrid synchrotron luminosity and
${\cal C}^{\rm nth}$ is the normalization constant for thermal Comptonization
spectrum of hybrid synchrotron photons.

To calculate the ratio \odn{e:softflux} we assume the optically thick
hybrid synchrotron emission below $\xnth$ to be $\propto S^{\rm pl}(x)$,
the source function of power-law electrons,
\begin{equation}
\label{e:source}
S^{\rm pl}(x)=\frac{1}{2\,3^{1/2}}\frac{G_2}{G_1} \frac{m_{\rm
e}c^3}{\lambda_{\rm C}^3} x_{\rm c}^{-{1 \over 2}}x^{5 \over 2},
\end{equation}
where
\begin{equation}
\label{e:g2}
G_2=\frac{\Gamma(\frac{5+p}{4})\Gamma(\frac{3p+19}{12})\Gamma(\frac{3p-1}{12})}
{\Gamma(\frac{7+p}{4})} \simeq 1.
\end{equation}
The emission above $\nnth$ is $\propto \nu^{-(p-1)/2}$ and then we have
\begin{equation}
\label{e:cnth}
\frac{{\cal C}^{\rm nth}}{{\cal C}^{\rm th}}=\frac{3^{1\over2}(p+4)G_2
\left(\xnth/x_{\rm c}\right)^{7/2}}{14(p-3)G_1 \Theta \left(\xtth/x_{\rm
c}\right)^{3}}.
\end{equation}
The above expressions were derived using the emission and absorption
coefficients for power-law electrons (e.g.\ Rybicki \& Lightman 1979). Since we
neglected the cut-off in the electron distribution, the above relation holds
for $\xnth/x_{\rm c}\ll\gamma_{\rm f}^2$, i.e.\ when the contribution from
electrons with $\gamma>\gamma_{\rm f}$ to the emission at $\xnth$ can be
neglected.

The total luminosity of the CS process in the hybrid case is then
\begin{equation}
\label{e:lcsnth}
L^{\rm nth}_{\rm CS}=L_{\rm S}^{\rm nth} + L^{\rm th}_{\rm C}({\cal C}^{\rm
nth})+L^{\rm pl}_{\rm C}({\cal C}^{\rm nth}).
\end{equation}
When $\alpha<1$, we can neglect $L_{\rm S}^{\rm th}$ and $L_{\rm S}^{\rm nth}$
in equations \odn{e:lcsth} and \odn{e:lcsnth} respectively, since most of the
luminosity is then produced at much higher energies, $x \sim \Theta$. Then, the
ratio $L_{\rm CS}^{\rm nth}/L_{\rm CS}^{\rm th}$ can be calculated as
\begin{eqnarray}
\label{e:csratio}
\lefteqn{\frac{L_{\rm CS}^{\rm nth}}{L_{\rm CS}^{\rm th}} = \frac{{\cal
C}^{\rm nth}}{{\cal C}^{\rm th}} \left[1+\frac{L_{\rm C}^{\rm pl}\left({\cal
C}^{\rm th}\right)}{L_{\rm C}^{\rm th}\left({\cal C}^{\rm th}\right)}\right].}
\end{eqnarray}
We can neglect the last term in the brackets if $\delta$ is sufficiently
small, in which case $L_{\rm C}^{\rm pl}\left({\cal C}^{\rm th}\right) \ll
L_{\rm C}^{\rm th}\left({\cal C}^{\rm th}\right)$. We then have
\begin{eqnarray}
\label{e:apprcsratio}
\lefteqn{\frac{L_{\rm CS}^{\rm nth}}{L_{\rm CS}^{\rm th}} \approx
\frac{3^{1\over2}(p+4)\left(\xnth/x_{\rm c}\right)^{7/2}}{14(p-3) \Theta
\left(\xtth/x_{\rm c}\right)^{3}}.}
\end{eqnarray}

Note that in calculations of ${\cal C}^{\rm nth}$ we used the source function
for power-law electrons. Thus, this approximation does not have the correct
thermal limit when $\gamma_{\rm nth} \gg \gamma_{\rm t}$, i.e. when $x_{\rm
t}^{\rm nth} \approx x_{\rm t}^{\rm th}$ (see Fig.\ \ref{f:widma}b, where a
synchrotron spectrum for such a case is shown). Note also that for harder
power-law electron distributions ($p\la3$), for which the approximation
\odn{e:softflux} does not hold, the amplification of CS luminosity will be even
larger than that predicted above.

\subsection{Applications to NGC 4151, GX 339--4 and Cyg X-1}
\label{s:applications}

We now calculate the ratio $L_{\rm CS}^{\rm nth}/L_{\rm CS}^{\rm th}$ as a
function of the energy content of the electron power-law tail in three
different cases. They correspond to the models of thermal CS emission from a
hot accretion disc with magnetic field in equipartition with the gas internal
energy (dominated by hot ions) of WZ00 for the Seyfert galaxy NGC 4151 and two
BHBs in their hard states, GX 339--4 and Cyg X-1. The CS luminosities predicted
by the thermal model (fig.\ 8a in WZ00) were orders of magnitude below the ones
observed  from NGC 4151 and GX 339--4, and less by a factor of $\sim 3$ for
Cyg~X-1.

\begin{figure} \begin{center} \noindent\epsfxsize=8.4cm\epsfbox{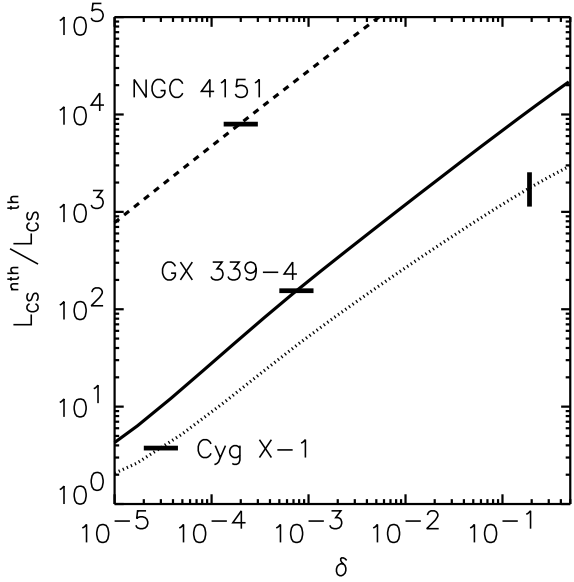}
\end{center}
\caption{The amplification of the CS emission due to the presence of nonthermal
electrons, $L_{\rm CS}^{\rm nth}/L_{\rm CS}^{\rm th}(\delta)$, for models (see
text) of NGC 4151, GX 339--4 and Cyg X-1 (dashed, solid and dotted curves,
respectively). The horizontal lines correspond to the amplification required to
account for the observed X-ray luminosities and the corresponding values of
$\gamma_{\rm nth}$ are 2.61, 2.46, and 4.10 in the above 3 objects,
respectively. The vertical line corresponds to $\delta$ obtained from
observations of the non-thermal tail in
the spectrum of Cyg X-1 by McConnell et al.\ (2000a), which corresponds to the
X-ray luminosity exceeding that observed by a factor of $\sim 500$.
}
\label{f:powers}
\end{figure}

For NGC 4151, we assume (see WZ00 for details and references) $R=1.2 \times
10^{14}$ cm, $B=3.8 \times 10^3$ G, $\alpha=0.85$, $\Theta=0.1$, $p=4$,
$\gamma_{\rm f}=100$. For GX~339--4, the same values of $\Theta$, $p$ and
$\gamma_{\rm f}$ are adopted while $R=8.8 \times 10^{6}$ cm, $B=1.5 \times
10^7$ G, $\alpha=0.75$. In the case of Cyg X-1, $R=3 \times 10^{7}$ cm, $B=6.3
\times 10^6$ G, $\alpha=0.6$, and we use the electron distribution obtained in
McConnell et al. (2000a) by fitting the spectra from the COMPTEL and OSSE
detectors aboard {\it CGRO}, i.e.\ $\Theta=0.17$, $\gamma_{\rm nth}=2.12$,
$p=4.5$, at assumed $\gamma_{\rm f} =10^3$, which corresponds to a rather large
value of $\delta=0.19$.

The results are shown in Fig.\ \ref{f:powers}. In the case of NGC~4151 and GX
339--4, a non-thermal tail in the electron distribution with $\delta \la
10^{-3}$ amplifies the power produced by the CS process sufficiently to
reproduce the luminosities of those sources. Therefore, unlike the results
obtained by WZ00 from investigating the purely thermal CS process, in the case
of electron distributions with relatively weak non-thermal tails the CS process
appears to be capable of producing the observed X-ray spectra of accreting
black holes, independent of the Eddington ratio or the black hole mass.

On the other hand, the estimated {\em thermal} CS luminosity in the case of Cyg
X-1 in the hot disc model is comparable to that observed from the source. Then,
we find that the non-thermal tail observed by McConnell et al. (2000a) would
produce the CS luminosity $\sim 500$ times higher than that actually observed
(see Fig. \ref{f:widma}a, where thermal and hybrid synchrotron spectra are
plotted for plasma parameters assumed in this model). The requirement that
$L_{\rm CS}^{\rm nth}$ cannot exceed the source luminosity constrains the
magnetic field strength in the model to be $B\la 2.3 \times 10^5$ G, which is
$\sim 30$ times smaller than the equipartition value estimated by WZ00.

A similar constraint for Cyg X-1 can be obtained in a model of active coronal
regions above the disc with magnetic field dissipation (see WZ00 for details of
the corresponding model of CS emission). Assuming $N=10$ active regions of
radius $R_{\rm b}=3 \times 10^6$ cm and height $H_{\rm b}=3 \times 10^5$ cm, we
obtain $B\la 3.4 \times 10^6$ G, much less than the field strength predicted
assuming dissipation of magnetic field at a fraction of the Alfv\'en speed,
$B_{\rm diss}=5 \times 10^7$ G. Though varying  $N$, $R_{\rm b}$, $H_{\rm b}$
would change the upper limit on the field strength as well as the value of
$B_{\rm diss}$, the relation between these quantities would remain similar.
This suggests that the mechanism of dissipation of disc field in small active
regions above the disc cannot be responsible for the energy release in the
corona. This stems from the fact that $B_{\rm diss}$ is the minimum field
strength necessary for the magnetic field to dissipate enough energy so as to
produce the observed luminosity. Therefore, either the model of active regions
above a cold disc is ruled out for the hard state of Cyg X-1 or another
mechanism of energy release must operate. We point out that if the CS process
were to be negligible (as suggested by observations, see Section
\ref{s:discussion}), the magnetic field would have to be even weaker, which
would strengthen our conclusion even more. On the other hand, a weak magnetic
field does not present analogous difficulties in the hot accretion disc model
as the disc heating is directly by gravity.

We then check how sensitive the above results are to details of the observed
spectra of Cyg X-1. First, the spectral index of Cyg X-1 varies, and the value
$\alpha=0.6$ assumed above is the hardest one and spectra with $\alpha=0.7$
have been also observed. Second, the relative normalization between the COMPTEL
and OSSE spectra remains relatively uncertain, and it may be possible that the
former is less by $\sim 2$ (McConnell et al.\ 2000b) than the one used in the
calculations above. This would then diminish the energy content in the electron
non-thermal tail twice, which corresponds to the new value of $\gamma_{\rm
nth}=2.31$. Thus, we have repeated our calculations for $\alpha=0.7$ and
$\gamma_{\rm nth}=2.31$. The resulting constraints on the magnetic field
strength in the hot disc and active regions models are then $B\la 5.8 \times
10^5$ G and $B \la 6.8 \times 10^6$ G, respectively. These values are only
about twice the previous upper limits, which does not affect our conclusions
above.

We note that another constraint on the CS emission may be provided by
extrapolating the X-ray power law spectrum to lower energies (but $\ga
h\nu_{\rm t}$), e.g.\ to the UV or optical ranges. That emission should not,
in any case, exceed that observed. In the case of Cyg X-1, we have extrapolated
the observed X-ray power-law spectrum of Cyg X-1 (Gierli\'nski et al. 1997) to
$\nu \simeq 10^{15}$ Hz and have found it is more than 2 orders of magnitude
below the UV spectrum of the companion supergiant as observed by {\it IUE\/}
(Treves et al.\ 1980). Thus, those data are compatible with the CS origin of
the X-ray spectrum in Cyg X-1.

On the other hand, GX 339--4 is a low-mass X-ray binary with a low upper limit
on the emission of the companion star (e.g.\ Zdziarski et al.\ 1998). Then,
simultaneous (as the object is highly variable, e.g.\ Corbet et al.\ 1987)
observations in X-rays and at longer wavelengths can put constraints on the
CS model in this case (e.g.\ Fabian et al.\ 1982).

\section{Conclusions and discussion}
\label{s:discussion}

We have studied the influence of non-thermal tails in the electron distribution
on the Comptonized synchrotron emission from plasmas of parameters typical for
accreting black holes. In those plasmas, most of the self-absorbed thermal
synchrotron emission is produced by electrons far in the tail of the Maxwellian
distribution. As a result, even a weak non-thermal component far beyond the
thermal peak can greatly enhance the synchrotron emission. This emission
provides seed photons for Comptonization, which is also correspondingly
enhanced. However, the shape the Comptonization spectrum remains only weakly
modified, with the main signature being a relatively weak high-energy tail.

We have shown that this effect becomes stronger with the decreasing plasma
temperature and increasing optical depth, which typically correspond to an
increasing Eddington ratio. It is also much stronger in the case of AGNs than
of accreting stellar black holes, as a result of weaker magnetic field in the
former objects. The CS luminosity can be amplified by a factor $\sim 10^5$ and
$\sim 10^3$, respectively, in the case of only $\sim 1$ per cent of the
electron energy in the non-thermal tail.

Then, the Comptonized synchrotron emission becomes energetically capable of
producing the observed luminosities of accreting black holes independently of
their Eddington ratio or the black-hole mass. This conclusion is qualitatively
different from that obtained in the case of purely thermal plasma.

It is then possible to obtain upper limits on the magnetic field strength in
optically thin plasmas around black holes by constraining the parameters of the
non-thermal component in the electron distribution from observations of the
high-energy tail in Comptonization spectra. Based on the {\it CGRO\/} data for
Cyg X-1 in the hard state, we have obtained the upper limits on the field
strength at least an order of magnitude below both the value corresponding to
equipartition (in the model of a two-temperature accretion disc) and the
minimum value required by dissipation of magnetic fields (in the model of
active coronal regions). Therefore, the latter model appears to be ruled out
for the hard state of Cyg X-1.

We note that the presence of a strong correlation between reflection strength
and the X-ray spectral index, observed in Cyg X-1 (Gilfanov et al.\ 1999),
suggests that the CS emission is a negligible source of photons for
Comptonization and thus the magnetic field strength must remain significantly
below the upper limits derived above. The same correlation seen in other
objects (Zdziarski et al.\ 1999; Gilfanov et al.\ 2000) implies that either the
high-energy tails or the magnetic fields are weak enough for the CS process not
to dominate the energy output.

In this work, we have assumed, for the sake of simplicity and compatibility
with studies of, e.g., McConnell et al.\ (2000a), the electron distribution of
the non-thermal electron tails to be $\propto \gamma^{-p}$ at any value of
$\gamma$. However, when the power-law tail extends down to non-relativistic
energies, $\gamma_{\rm nth} \sim 1$, the distributions expected from
acceleration processes are power laws in either the kinetic energy or the
momentum, e.g.\ $\propto (\gamma-1)^{- p}$ for the former. Clearly, the smaller
$\gamma_{\rm nth}$ (i.e.\ the larger $\delta$), the larger the differences
between those distributions. Then, the actual shift of the turnover frequency
at a given value of $\gamma_{\rm nth}$ would be somehow smaller. Also, the
relation between $\gamma_{\rm nth}$ and $\delta$ would change. Nevertheless,
the above effects would modify our results only quantitatively without
affecting our conclusions.

The influence of a weak, non-thermal, component in the electron distribution on
radiation spectra has also been independently considered by \"Ozel, Psaltis \&
Narayan (2000). Their study has been devoted to the case of
advection-dominated accretion flows (ADAF), and their presented spectra are
integrated over all radii of the flow. In contrast to our results, they do not
note any shift in the turnover frequency at the peak of the synchrotron
spectrum. This appears to result from their study being constrained to the ADAF
model at low accretion rates, in which case electrons in the innermost parts of
the accretion flow reach rather high temperatures (at which even thermal
electrons reach relatively high Lorentz factors). Those temperatures are
significantly higher than those we consider based on observational data from
luminous accreting black holes. On the other hand, \"Ozel et al. (2000) find a
significant excess non-thermal emission at frequencies well below the peak of
the integrated synchrotron spectrum. This excess results from radiation emitted
at large radii, where the electron temperature is low, and where the
synchrotron emission (both optically thick and optically thin) around the local
turnover frequency can be strongly amplified by the presence of an electron
tail, as illustrated in Fig.\ \ref{f:widma} above.

\section*{ACKNOWLEDGMENTS}

This research has been supported in part by the Foundation for Polish Science
and KBN grants 2P03D01619 and 2P03D00624. We thank Marek Gierli\'nski for
providing us with his Monte-Carlo Comptonization code and for help with its
use. We are also grateful to Andrei Beloborodov, Joanna Miko{\l}ajewska and
Juri Poutanen for valuable discussions.

\bsp

\label{lastpage}

\end{document}